\documentclass[aps,pra,reprint,superscriptaddress]{revtex4-2}
\usepackage{amsmath}
\usepackage{amssymb}
\usepackage{braket}
\usepackage{physics}
\usepackage{mathtools}
\usepackage[caption=false]{subfig}
\usepackage{siunitx}

\begin{document}

\title{Single-shot single-mode optical two-parameter displacement estimation beyond classical limit}

\author{Fumiya Hanamura}
\affiliation{Department of Applied Physics, School of Engineering, The University of Tokyo, 7-3-1 Hongo, Bunkyo-ku, Tokyo 113-8656, Japan}
\author{Warit Asavanant}
\affiliation{Department of Applied Physics, School of Engineering, The University of Tokyo, 7-3-1 Hongo, Bunkyo-ku, Tokyo 113-8656, Japan}
\affiliation{Optical Quantum Computing Research Team, RIKEN center for Quantum Computing, 2-1 Hirosawa, Wako, Saitama 351-0198, Japan}
\author{Seigo Kikura}
\affiliation{Department of Applied Physics, School of Engineering, The University of Tokyo, 7-3-1 Hongo, Bunkyo-ku, Tokyo 113-8656, Japan}
\author{Moeto Mishima}
\affiliation{Department of Applied Physics, School of Engineering, The University of Tokyo, 7-3-1 Hongo, Bunkyo-ku, Tokyo 113-8656, Japan}
\author{Shigehito Miki}
\affiliation{Advanced ICT Research Institute, National Institute of Information and Communications Technology, 588-2 Iwaoka, Nishi-ku, Kobe, Hyogo 651-2492, Japan}
\affiliation{Graduate School of Engineering, Kobe University, 1-1 Rokkodai-cho, Nada-ku, Kobe, Hyogo 657-0013, Japan}
\author{Hirotaka Terai}
\affiliation{Advanced ICT Research Institute, National Institute of Information and Communications Technology, 588-2 Iwaoka, Nishi-ku, Kobe, Hyogo 651-2492, Japan}
\author{Masahiro Yabuno}
\affiliation{Advanced ICT Research Institute, National Institute of Information and Communications Technology, 588-2 Iwaoka, Nishi-ku, Kobe, Hyogo 651-2492, Japan}
\author{Fumihiro China}
\affiliation{Advanced ICT Research Institute, National Institute of Information and Communications Technology, 588-2 Iwaoka, Nishi-ku, Kobe, Hyogo 651-2492, Japan}
\author{Kosuke Fukui}
\affiliation{Department of Applied Physics, School of Engineering, The University of Tokyo, 7-3-1 Hongo, Bunkyo-ku, Tokyo 113-8656, Japan}
\author{Mamoru Endo}
\affiliation{Department of Applied Physics, School of Engineering, The University of Tokyo, 7-3-1 Hongo, Bunkyo-ku, Tokyo 113-8656, Japan}
\affiliation{Optical Quantum Computing Research Team, RIKEN center for Quantum Computing, 2-1 Hirosawa, Wako, Saitama 351-0198, Japan}
\author{Akira Furusawa}
\affiliation{Department of Applied Physics, School of Engineering, The University of Tokyo, 7-3-1 Hongo, Bunkyo-ku, Tokyo 113-8656, Japan}
\affiliation{Optical Quantum Computing Research Team, RIKEN center for Quantum Computing, 2-1 Hirosawa, Wako, Saitama 351-0198, Japan}

\date{\today}

\begin{abstract}
Uncertainty principle prohibits the precise measurement of both components of displacement parameters in phase space. We have theoretically shown that this limit can be beaten using single-photon states, in a single-shot and single-mode setting [F. Hanamura et al., Phys. Rev. A 104, 062601 (2021)]. In this paper, we validate this by experimentally beating the classical limit. In optics, this is the first experiment to estimate both parameters of displacement using non-Gaussian states. This result is related to many important applications, such as quantum error correction.
\end{abstract}


\maketitle

\textit{Introduction.-}
Uncertainty, which prohibits the simultaneous measurement of noncommutative observables $\hat{x}$ and $\hat{p}$ such that $[\hat{x},\hat{p}]=i$, is a fundamental feature of quantum mechanics. As far as classical states are used, uncertainty imposes limits to the precision of quantum sensors, which are called standard quantum limits (SQL) \cite{sql_metrology}, or classical limits. However, it has long been known that these limits can be circumvented using non-classical quantum states and measurements, and beating the classical limits has been an important topic for various applications such as gravitational wave detection \cite{gravitational_wave}, optical clock \cite{atomic_clock}, and optomechanical systems \cite{opto_mech}.

Displacement operation is a parallel translation operation in the phase space: $\hat{x}\to \hat{x}+\xi$, $\hat{p}\to \hat{p}+\eta$. The problem to estimate both components of unknown displacement parameters $\xi,\eta$ is a fundamental problem which is directly related to the uncertainty of $\hat{x}$ and $\hat{p}$. This problem becomes trivial in the case when the displacements with the same parameter are applied multiple times to an ensemble of quantum states, or the case of entangled states where the displacement is implemented on only single mode. In such cases, there are ways to precisely know $\xi,\eta$ using one-mode or two-mode squeezed states \cite{opt_est_joint}.

Thus the most interesting and non-trivial case is the single-mode and single-shot estimation of both parameters. Reference \cite{single_mode_disp_sensor} has shown that even in this situation, both parameters can be precisely known using so-called Gottesman-Kitaev-Preskill (GKP) states which has been originally proposed for quantum error correction \cite{gkp}. However, GKP state is a highly non-Gaussian state and its generation in optical system has not been achieved yet, although it has been generated in other physical platforms, such as superconducting cavity \cite{cqed_gkp}, and ion trap \cite{ion_gkp}. Thus, although there are some experimental attempts to use non-Gaussian states for quantum parameter estimation in both optics and other bosonic systems \cite{noon_optics,fock_estimation_ion,Radim_nongauss}, single-shot single-mode estimation of both parameters of displacements has not been experimentally demonstrated in optics. To deal with this problem, in Ref.~\cite{disp_estimation_theory} we have theoretically investigated the estimation of displacement parameters using exprimentally more feasible non-Gaussian states, and have shown that even single-photon states can beat the classical limit, with the newly introduced criteria to evaluate the estimation error using the variance of the posterior distribution after the post-selection of the measurement outcome.

In this paper, we experimentally demonstrate estimation of displacement parameters using single-photon states with accuracy beyond the classical limit. This is the first experimental demonstration of single-shot single-mode estimation of two parameters of displacement using non-Gaussian states in optical systems. Our result also can be considered as the first observation of the ability to sense displacements coming from the sub-Planck structure \cite{subPlanck,single_mode_disp_sensor} of optical quantum state. This result can serve as a fundamental result not only for quantum parameter estimation using non-Gaussian states, but also for the research of quantum error correction, as the single-shot single-mode estimation of both parameters of displacement is closely related to quantum error correction of Gaussian errors \cite{general_tscode,gkp,qec_metrology}.

\textit{Displacement estimation problem.-}
We will briefly explain the proposal in Ref.~\cite{disp_estimation_theory}. Displacement operator $\hat{D}(\xi,\eta)$ is defined as a translation in phase space:
\begin{align}
    \hat{D}^\dagger(\xi,\eta)\hat{x}\hat{D}(\xi,\eta)&=\hat{x}+\xi,\\
    \hat{D}^\dagger(\xi,\eta)\hat{p}\hat{D}(\xi,\eta)&=\hat{p}+\eta.
\end{align}
We consider a problem to estimate both parameters $\xi$ and $\eta$ in a single-shot experiment. The input state is restricted to single-mode states. We can perform post-selection of the measurement outcome, which may make the estimation error smaller in the expense of lower success probability.
\begin{figure}[h]
    \centering
    \includegraphics[width=0.9\linewidth]{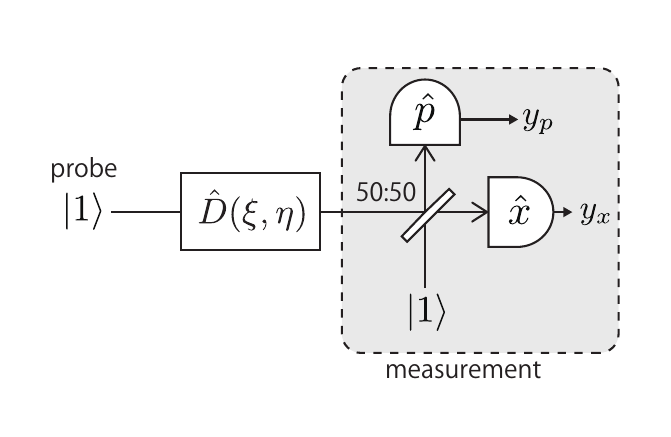}
    \caption{A schematic figure of the system to estimate displacements. After the displacement is applied to the input probe single-photon state, dual-homodyne measurement with an ancillary single-photon state is performed, obtaining the measurement outcome $y_x,y_p$.}
    \label{fig:displacement_estimation}
\end{figure}

\begin{figure}[h]
    \centering
    \includegraphics[width=0.9\linewidth]{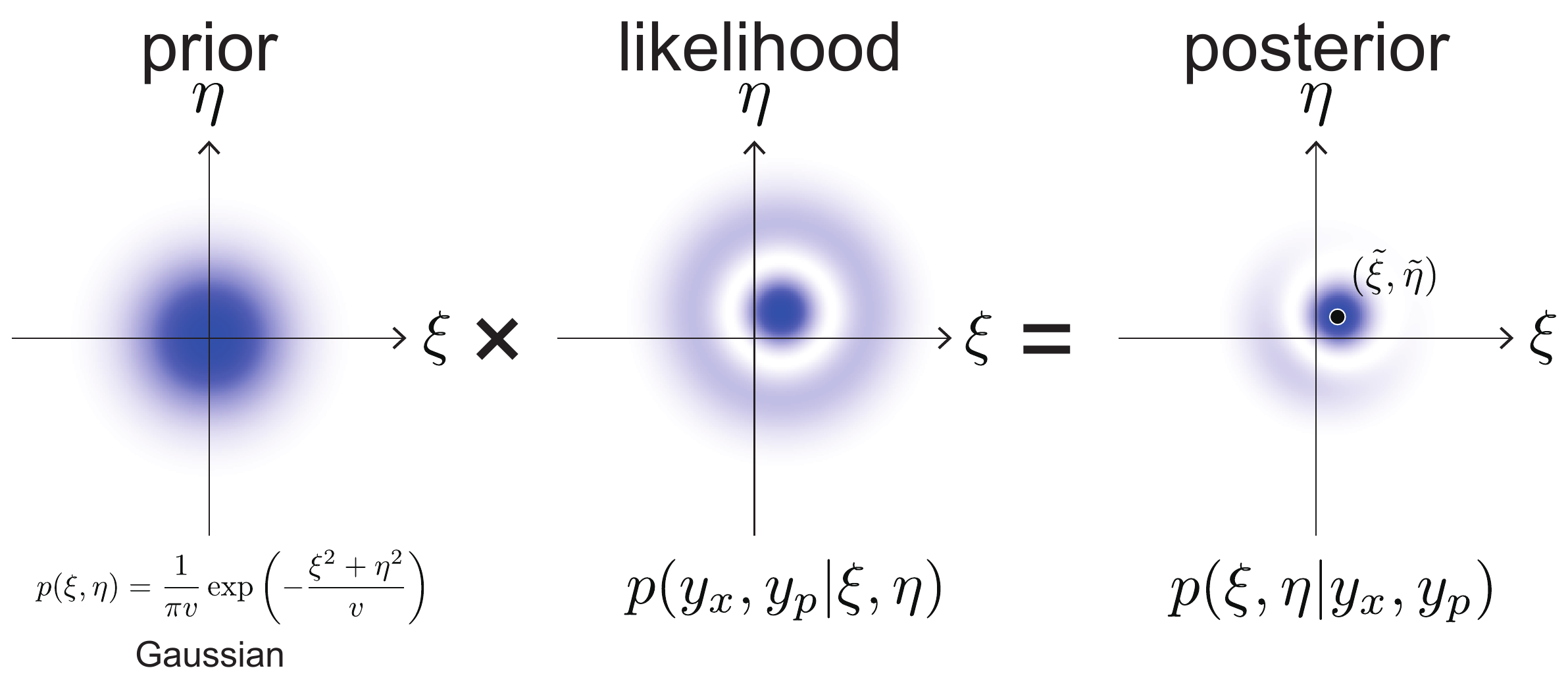}
    \caption{Bayesian estimation of displacement. The posterior distribution of $\xi,\eta$ after obtaining the measurement outcome is expressed as the product of the prior distribution and the likelihood function.}
    \label{fig:bayesian_est}
\end{figure}

The system in Fig.~\ref{fig:displacement_estimation} is proposed for the estimation of displacements. Two single-photon states are prepared, one as the probe to be input to the displacement operation, and the other as an ancillary state for the measurement. We consider the case where we have prior information that the distribution of $\xi,\eta$ is an isotropic Gaussian function
\begin{align}
    p(\xi,\eta)=\frac{1}{\pi v}\exp(-\frac{\xi^2+\eta^2}{v}),\label{eq:prior_dist}
\end{align}
which means we consider so-called Gaussian quantum channel or additive Gaussian noise \cite{gaussian_quantum_channel}.
When a measurement outcome $y_x,y_p$ is obtained, the posterior distribution of $\xi,\eta$ is expressed as the product of the prior distribution and the likelihood function
\begin{align}
    p(\xi,\eta|y_x,y_p)\propto p(\xi,\eta)p(y_x,y_p|\xi,\eta),
\end{align}
as shown in Fig.~\ref{fig:bayesian_est}.

The likelihood function $p(y_x,y_p|\xi,\eta)$ reflects the information of the input probe state and the measurement. In the case we consider, it is expressed as
\begin{align}
    p(y_x,y_p|\xi,\eta)\propto \qty(W_{1}*W_{2})\qty(\sqrt{2}y_x-\xi,\sqrt{2}y_p-\eta),\label{eq:likelihood}
\end{align}
where $W_1$ and $W_2$ are Wigner functions of the input probe single-photon state and the ancillary single-photon state for the measurement, and $*$ represents convolution. This expression holds for arbitrary input probe states and measurement ancillae, including imperfect single-photon states in experiment. The estimated value $\tilde{\xi}, \tilde{\eta}$ is calculated as the mean of the posterior distribution:
\begin{align}
    \tilde{\xi}&=\int\int \xi p(\xi,\eta|y_x,y_p)d\xi d\eta\\
    \tilde{\eta}&=\int\int \eta p(\xi,\eta|y_x,y_p)d\xi d\eta
\end{align}
The error of estimation $v'$ is defined as the sum of mean-square errors of $\xi$ and $\eta$:
\begin{align}
    v'&=\expval{(\xi-\tilde{\xi})^2}+\expval{(\eta-\tilde{\eta})^2}\label{eq:est_error}
\end{align}
where the average is taken over all post-selected events.

In Ref.~\cite{disp_estimation_theory}, we derived the Gaussian bound and the classical bound of the estimation error $v'$. We have shown that with post-selection
\begin{align}
    y_x^2+y_p^2<r^2\label{eq:homodyne_postselection}
\end{align}
for small value of $r$, pure single-photon states can beat the Gaussian limit, and single-photon states up to 50\% loss can beat the classical limit. In the following sections, we show experimental results demonstrating the estimation error smaller than the classical limit.

\textit{Experimental setup.-}
\begin{figure*}[ht]
    \centering
    \includegraphics[width=0.8\linewidth]{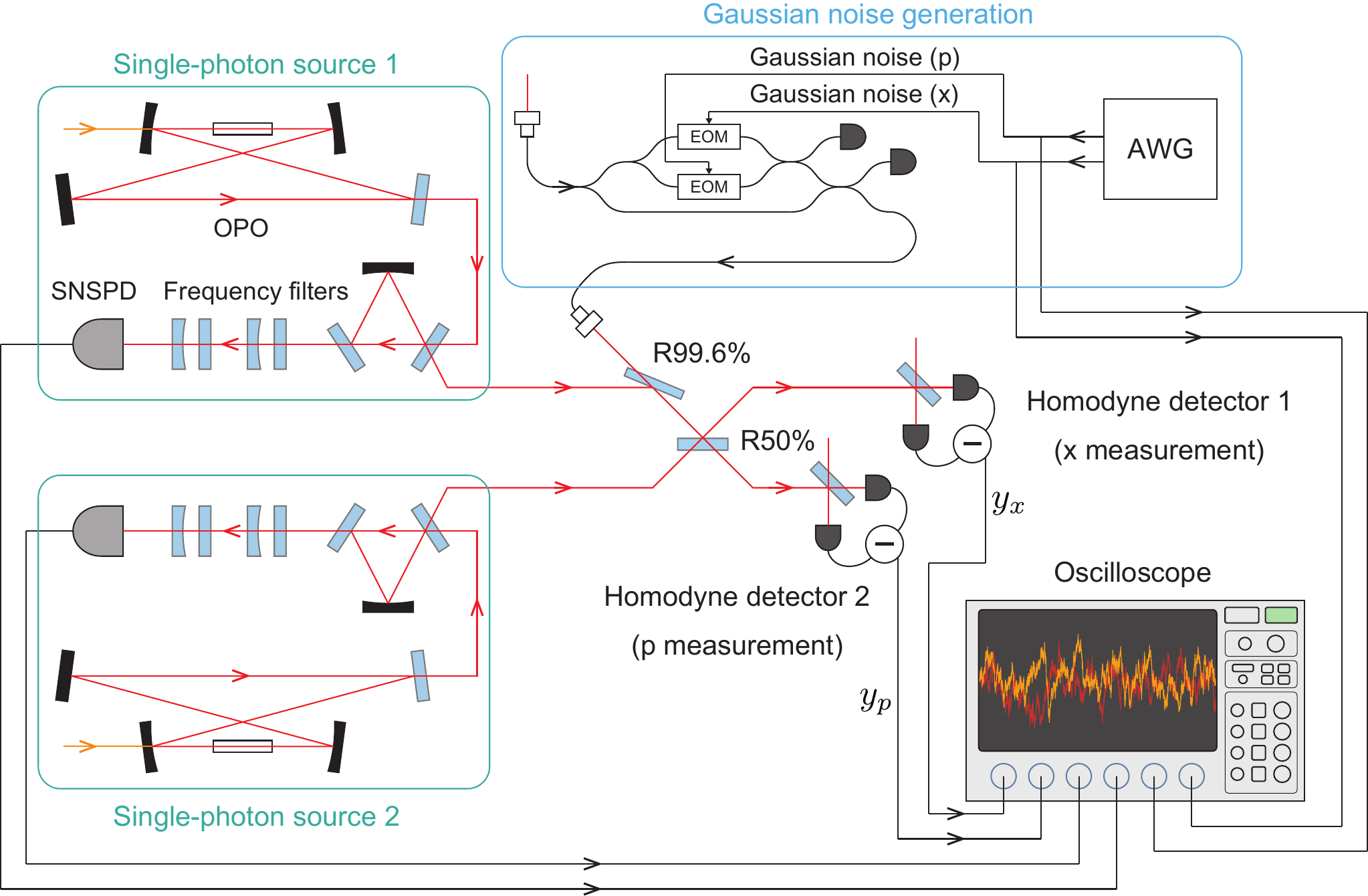}
    \caption{Experimental setup. OPO, optical parametric oscillator; SNSPD, superconducting nanostrip single-photon detector; EOM, electro-optic modulator; AWG, arbitrary waveform generator.}
    \label{fig:experimental_setup}
\end{figure*}
Figure \ref{fig:experimental_setup} shows our experimental setup. We use a continuous-wave laser with a wave length of 1545.32 nm. We prepare two independent single-photon sources, with heralding scheme using optical parametric oscillators (OPOs) with periodically poled KTiOPO4 (PPKTP) crystals of Type-0, and superconducting nanostrip single-photon detectors (SNSPDs) made of NbTiN \cite{sspd}. The FSR of the OPOs is \SI{1.1}{GHz} and the FWHM is \SI{25}{MHz}. The event rate is around \SI{1e4}{cps} for each single-photon source. Photon coincidence events where two trigger timings $t_1$ and $t_2$ satisfies $|t_1-t_2|<\SI{5}{ns}$ are collected.

Random coherent states whose complex amplitude follow isotropic Gaussian distribution Eq.~(\ref{eq:prior_dist}) with variable $v$ is generated using waveguide electro-optic modulators (EOMs) and a fiber interferometer. The frequency bandwidth of the Gaussian noise is \SI{6}{MHz}. Using the random coherent state, we apply random displacement to one of the single-photon state. After interfering at the 50:50 beamsplitter, $x$ and $p$ quadratures are measured by homodyne detectors respectively on two output modes, obtaining the measurement outcomes $y_x$ and $y_p$. 

\textit{Results.-}
\begin{figure}[hbt]
    \centering
    \subfloat[]{\includegraphics[width=0.45\linewidth]{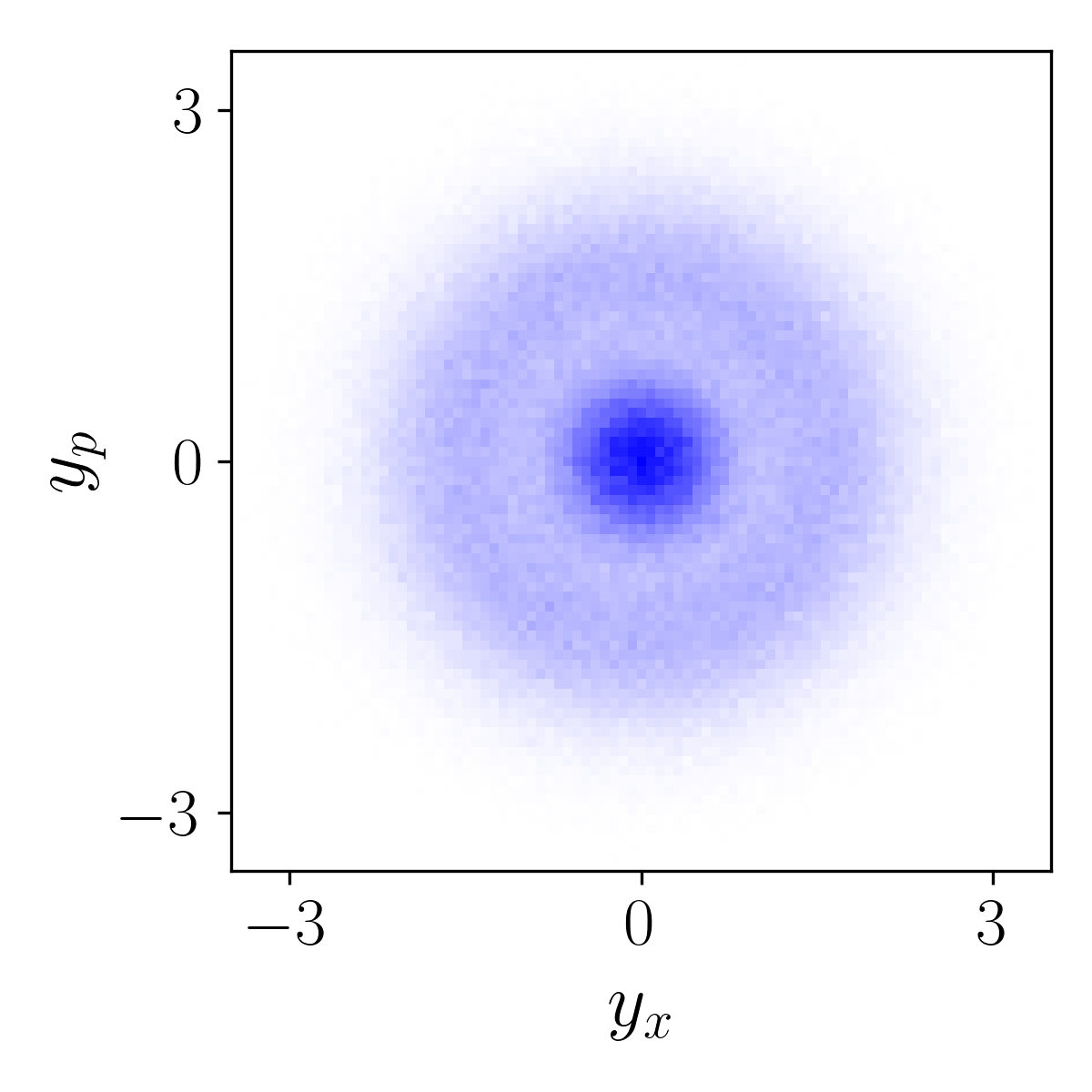}\label{fig:likelihood_2d}}
    \subfloat[]{\includegraphics[width=0.55\linewidth]{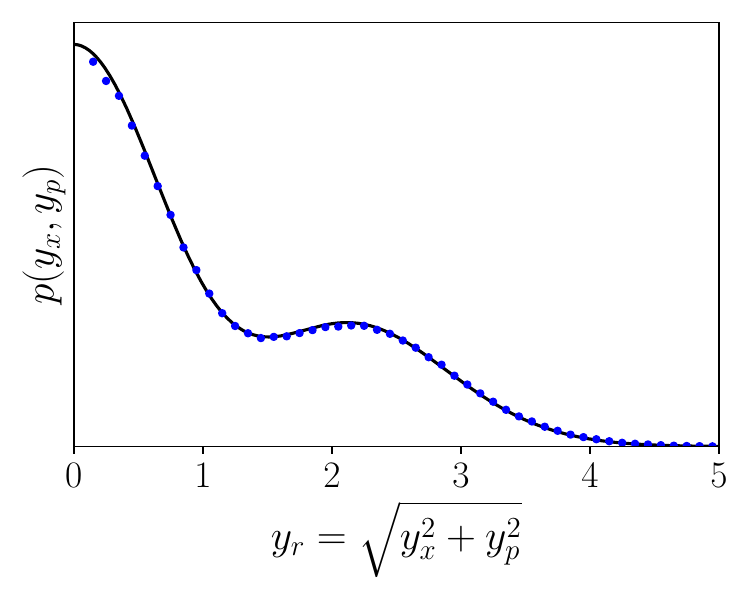}\label{fig:likelihood_1d}}\\
    \subfloat[]{\includegraphics[width=0.5\linewidth]{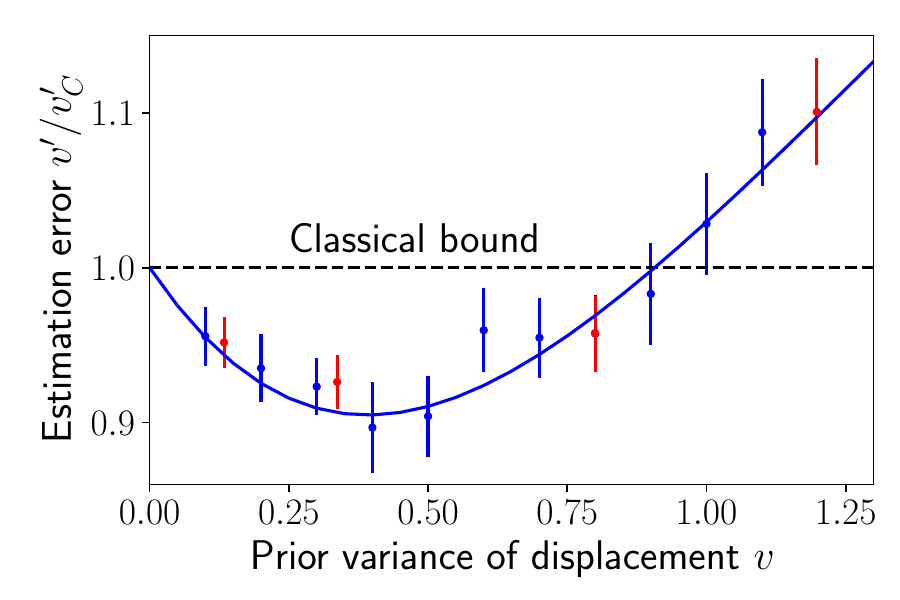}\label{fig:v_vs_vp}}
    \subfloat[]{\includegraphics[width=0.5\linewidth]{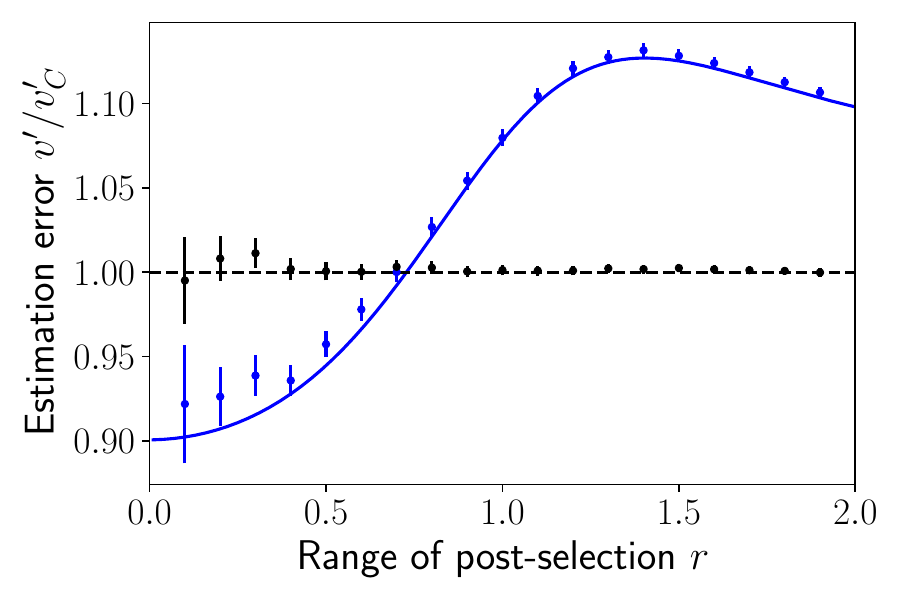}\label{fig:select_prob_vs_vp}}
    \caption[]{Experimental results. \subref{fig:likelihood_2d} The simultaneous distribution of measurement outcomes $y_x,y_p$ when no displacement is applied. \subref{fig:likelihood_1d} Blue dot: Radial profile of the distribution of \subref{fig:likelihood_2d}. Black solid line: Theoretical prediction when imperfect single-photon states with 25\% 0-photon component and 2\% 2-photon component are used. \subref{fig:v_vs_vp} Estimation error $v'$ normalized by the classical limit $v'_C$ as a function of the prior variance of displacement $v$, with $r=0.2$. $v$ corresponding to the red points are actually measured, and the blue points are calculated by post-selecting the events so that the prior distribution of displacements has the desired $v$. A theoretical curve for the same assumption as \subref{fig:likelihood_1d} is shown together. \subref{fig:select_prob_vs_vp} Estimation error $v'$ normalized by the classical limit $v'_C$ as a function of the range of post-selection $r$, with $v=0.34$. A theoretical curve for the same assumption as \subref{fig:likelihood_1d} is shown together.}
    \label{fig:vp_plots}
\end{figure}
 We first set $v=0$ (no displacement), and measured the simultaneous distribution $p(y_x,y_p|0,0)$ of homodyne measurement outcomes (Figs.~\ref{fig:likelihood_2d} and \ref{fig:likelihood_1d}). This matches well with the theoretical prediction when imperfect single-photon states with 25\% vacuum and 2\% two-photon components are used (solid line in Fig.~\ref{fig:likelihood_1d}). Note that from Eq.~(\ref{eq:likelihood}), this distribution gives the information about the shape of the likelihood function $p(\xi,\eta|y_x,y_p)$. Sharp peak near the origin represents the sub-Planck structure of the Wigner function of single-photon states which enables higher estimation accuracy. 
 
 Secondly, $v$ is set to $v=0.13,0.34,0.8,1.2$, and $165741, 168917, 125339, 117375$ events are collected for each case respectively. Estimated values $\tilde\xi, \tilde\eta$ of displacement parameters $\xi, \eta$ are calculated from the measurement outcomes $y_x,y_p$ and the estimated likelihood function $p(y_x,y_p|\xi,\eta)$. Events where the homodyne measurement outcomes satisfy Eq.~(\ref{eq:homodyne_postselection}) are collected, and the estimation error $v'$ (Eq.~(\ref{eq:est_error})) is evaluated. Figure \ref{fig:v_vs_vp} shows the relative estimation error $v'$ normalized by the classical limit, where post-selection range in Eq.~(\ref{eq:homodyne_postselection}) is taken as $r=0.2$. Red points represent actually measured data points. In order to increase the data points, different values of $v$ are simulated by post-selecting events so that the distribution of $\xi,\eta$ has the desired variances (blue points in Fig.~\ref{fig:v_vs_vp}). Theoretical prediction in the case of imperfect single-photon states is shown together (blue solid line). The classical limit is beaten in the range of $0<v<0.9$. For $v=0.34$, the dependence on the range of post-selection $r$ is also tested (Fig.~\ref{fig:select_prob_vs_vp}). The classical limit is beaten up to $r<0.7$. We also used vacuum inputs to verify the classical limit (black points in Fig.~\ref{fig:select_prob_vs_vp}), which matches quite well with the theoretically calculated classical limit.

\textit{Conclusion and discussion.-}
In summary, we have demonstrated single-mode, single-shot displacement estimation with accuracy beyond the classical limit using single-photon states, homodyne measurements, and post-selection. This is the first optical demonstration of the ability to sense both parameters of displacement which emerges from the sub-Planck feature of non-Gaussian states. This experiment can serve as a fundamental result for future researches in quantum parameter estimation using non-Gaussian states.

As a future work, this system can be extended to arbitrary input probe state and measurement ancilla, as mentioned in Ref. \cite{disp_estimation_theory}. Using states with more non-Gaussianity leads to the estimation precision better than Gaussian limit even without the post-selection.

It is also worth noting that the setting considered here is also closely related to the quantum error correction. For example, GKP error correction \cite{gkp} relies on the ability of the GKP state to detect small displacement errors, especially Gaussian random displacement errors just like in our setting. This also needs to be single-shot, single-mode estimation, as the displacement is random and acts on each mode independently. The difference here is that in quantum error correction, one needs to estimate the displacement without destroying the quantum information encoded, while in this paper we estimate the displacement by measuring all modes thus destroy the quantum information. The correspondence is more clear in the two-mode squeezing code \cite{ts_code,general_tscode}, which is based on classical correlation between the displacements on different modes. In this case, one can estimate the displacement by performing destructive measurement, although additional optimization of estimation process would be necessary because of the correlation between displacements. Investigating more concrete correspondence is an interesting topic for future researches.
    
The authors acknowledge supports from UTokyo Foundation and donations from Nichia Corporation of Japan. W.A. and M.E. acknowledge supports from Research Foundation for OptoScience and Technology. F.H. acknowledges supports from FoPM, WINGS Program, the University of Tokyo. This work was partly supported by JSPS KAKENHI (18H05207, 23KJ0498, 23K13040) and JST (Moonshot R\&D JPMJMS2064, JPMJMS2066, JPMJPR2254).

\bibliography{main.bib}

\begin{thebibliography}{19}%
\makeatletter
\providecommand \@ifxundefined [1]{%
 \@ifx{#1\undefined}
}%
\providecommand \@ifnum [1]{%
 \ifnum #1\expandafter \@firstoftwo
 \else \expandafter \@secondoftwo
 \fi
}%
\providecommand \@ifx [1]{%
 \ifx #1\expandafter \@firstoftwo
 \else \expandafter \@secondoftwo
 \fi
}%
\providecommand \natexlab [1]{#1}%
\providecommand \enquote  [1]{``#1''}%
\providecommand \bibnamefont  [1]{#1}%
\providecommand \bibfnamefont [1]{#1}%
\providecommand \citenamefont [1]{#1}%
\providecommand \href@noop [0]{\@secondoftwo}%
\providecommand \href [0]{\begingroup \@sanitize@url \@href}%
\providecommand \@href[1]{\@@startlink{#1}\@@href}%
\providecommand \@@href[1]{\endgroup#1\@@endlink}%
\providecommand \@sanitize@url [0]{\catcode `\\12\catcode `\$12\catcode
  `\&12\catcode `\#12\catcode `\^12\catcode `\_12\catcode `\%12\relax}%
\providecommand \@@startlink[1]{}%
\providecommand \@@endlink[0]{}%
\providecommand \url  [0]{\begingroup\@sanitize@url \@url }%
\providecommand \@url [1]{\endgroup\@href {#1}{\urlprefix }}%
\providecommand \urlprefix  [0]{URL }%
\providecommand \Eprint [0]{\href }%
\providecommand \doibase [0]{https://doi.org/}%
\providecommand \selectlanguage [0]{\@gobble}%
\providecommand \bibinfo  [0]{\@secondoftwo}%
\providecommand \bibfield  [0]{\@secondoftwo}%
\providecommand \translation [1]{[#1]}%
\providecommand \BibitemOpen [0]{}%
\providecommand \bibitemStop [0]{}%
\providecommand \bibitemNoStop [0]{.\EOS\space}%
\providecommand \EOS [0]{\spacefactor3000\relax}%
\providecommand \BibitemShut  [1]{\csname bibitem#1\endcsname}%
\let\auto@bib@innerbib\@empty
\bibitem [{\citenamefont {Caves}(1981)}]{sql_metrology}%
  \BibitemOpen
  \bibfield  {author} {\bibinfo {author} {\bibfnamefont {C.~M.}\ \bibnamefont
  {Caves}},\ }\bibfield  {title} {\bibinfo {title} {Quantum-mechanical noise in
  an interferometer},\ }\href {https://doi.org/10.1103/PhysRevD.23.1693}
  {\bibfield  {journal} {\bibinfo  {journal} {Phys. Rev. D}\ }\textbf {\bibinfo
  {volume} {23}},\ \bibinfo {pages} {1693} (\bibinfo {year}
  {1981})}\BibitemShut {NoStop}%
\bibitem [{\citenamefont {Abbott}\ \emph {et~al.}(2016)\citenamefont {Abbott},
  \citenamefont {Abbott}, \citenamefont {Abbott},\ and\ \citenamefont
  {Abernathy}}]{gravitational_wave}%
  \BibitemOpen
  \bibfield  {author} {\bibinfo {author} {\bibfnamefont {B.~P.}\ \bibnamefont
  {Abbott}}, \bibinfo {author} {\bibfnamefont {R.}~\bibnamefont {Abbott}},
  \bibinfo {author} {\bibfnamefont {T.~D.}\ \bibnamefont {Abbott}},\ and\
  \bibinfo {author} {\bibfnamefont {M.~R.}\ \bibnamefont {Abernathy}} (\bibinfo
  {collaboration} {LIGO Scientific Collaboration and Virgo Collaboration}),\
  }\bibfield  {title} {\bibinfo {title} {Observation of gravitational waves
  from a binary black hole merger},\ }\href
  {https://doi.org/10.1103/PhysRevLett.116.061102} {\bibfield  {journal}
  {\bibinfo  {journal} {Phys. Rev. Lett.}\ }\textbf {\bibinfo {volume} {116}},\
  \bibinfo {pages} {061102} (\bibinfo {year} {2016})}\BibitemShut {NoStop}%
\bibitem [{\citenamefont {Ludlow}\ \emph {et~al.}(2015)\citenamefont {Ludlow},
  \citenamefont {Boyd}, \citenamefont {Ye}, \citenamefont {Peik},\ and\
  \citenamefont {Schmidt}}]{atomic_clock}%
  \BibitemOpen
  \bibfield  {author} {\bibinfo {author} {\bibfnamefont {A.~D.}\ \bibnamefont
  {Ludlow}}, \bibinfo {author} {\bibfnamefont {M.~M.}\ \bibnamefont {Boyd}},
  \bibinfo {author} {\bibfnamefont {J.}~\bibnamefont {Ye}}, \bibinfo {author}
  {\bibfnamefont {E.}~\bibnamefont {Peik}},\ and\ \bibinfo {author}
  {\bibfnamefont {P.~O.}\ \bibnamefont {Schmidt}},\ }\bibfield  {title}
  {\bibinfo {title} {Optical atomic clocks},\ }\href
  {https://doi.org/10.1103/RevModPhys.87.637} {\bibfield  {journal} {\bibinfo
  {journal} {Rev. Mod. Phys.}\ }\textbf {\bibinfo {volume} {87}},\ \bibinfo
  {pages} {637} (\bibinfo {year} {2015})}\BibitemShut {NoStop}%
\bibitem [{\citenamefont {Aspelmeyer}\ \emph {et~al.}(2014)\citenamefont
  {Aspelmeyer}, \citenamefont {Kippenberg},\ and\ \citenamefont
  {Marquardt}}]{opto_mech}%
  \BibitemOpen
  \bibfield  {author} {\bibinfo {author} {\bibfnamefont {M.}~\bibnamefont
  {Aspelmeyer}}, \bibinfo {author} {\bibfnamefont {T.~J.}\ \bibnamefont
  {Kippenberg}},\ and\ \bibinfo {author} {\bibfnamefont {F.}~\bibnamefont
  {Marquardt}},\ }\bibfield  {title} {\bibinfo {title} {Cavity optomechanics},\
  }\href {https://doi.org/10.1103/RevModPhys.86.1391} {\bibfield  {journal}
  {\bibinfo  {journal} {Rev. Mod. Phys.}\ }\textbf {\bibinfo {volume} {86}},\
  \bibinfo {pages} {1391} (\bibinfo {year} {2014})}\BibitemShut {NoStop}%
\bibitem [{\citenamefont {Genoni}\ \emph {et~al.}(2013)\citenamefont {Genoni},
  \citenamefont {Paris}, \citenamefont {Adesso}, \citenamefont {Nha},
  \citenamefont {Knight},\ and\ \citenamefont {Kim}}]{opt_est_joint}%
  \BibitemOpen
  \bibfield  {author} {\bibinfo {author} {\bibfnamefont {M.~G.}\ \bibnamefont
  {Genoni}}, \bibinfo {author} {\bibfnamefont {M.~G.~A.}\ \bibnamefont
  {Paris}}, \bibinfo {author} {\bibfnamefont {G.}~\bibnamefont {Adesso}},
  \bibinfo {author} {\bibfnamefont {H.}~\bibnamefont {Nha}}, \bibinfo {author}
  {\bibfnamefont {P.~L.}\ \bibnamefont {Knight}},\ and\ \bibinfo {author}
  {\bibfnamefont {M.~S.}\ \bibnamefont {Kim}},\ }\bibfield  {title} {\bibinfo
  {title} {Optimal estimation of joint parameters in phase space},\ }\href
  {https://doi.org/10.1103/PhysRevA.87.012107} {\bibfield  {journal} {\bibinfo
  {journal} {Phys. Rev. A}\ }\textbf {\bibinfo {volume} {87}},\ \bibinfo
  {pages} {012107} (\bibinfo {year} {2013})}\BibitemShut {NoStop}%
\bibitem [{\citenamefont {Duivenvoorden}\ \emph {et~al.}(2017)\citenamefont
  {Duivenvoorden}, \citenamefont {Terhal},\ and\ \citenamefont
  {Weigand}}]{single_mode_disp_sensor}%
  \BibitemOpen
  \bibfield  {author} {\bibinfo {author} {\bibfnamefont {K.}~\bibnamefont
  {Duivenvoorden}}, \bibinfo {author} {\bibfnamefont {B.~M.}\ \bibnamefont
  {Terhal}},\ and\ \bibinfo {author} {\bibfnamefont {D.}~\bibnamefont
  {Weigand}},\ }\bibfield  {title} {\bibinfo {title} {Single-mode displacement
  sensor},\ }\href {https://doi.org/10.1103/PhysRevA.95.012305} {\bibfield
  {journal} {\bibinfo  {journal} {Phys. Rev. A}\ }\textbf {\bibinfo {volume}
  {95}},\ \bibinfo {pages} {012305} (\bibinfo {year} {2017})}\BibitemShut
  {NoStop}%
\bibitem [{\citenamefont {Gottesman}\ \emph {et~al.}(2001)\citenamefont
  {Gottesman}, \citenamefont {Kitaev},\ and\ \citenamefont {Preskill}}]{gkp}%
  \BibitemOpen
  \bibfield  {author} {\bibinfo {author} {\bibfnamefont {D.}~\bibnamefont
  {Gottesman}}, \bibinfo {author} {\bibfnamefont {A.}~\bibnamefont {Kitaev}},\
  and\ \bibinfo {author} {\bibfnamefont {J.}~\bibnamefont {Preskill}},\
  }\bibfield  {title} {\bibinfo {title} {Encoding a qubit in an oscillator},\
  }\href {https://doi.org/10.1103/PhysRevA.64.012310} {\bibfield  {journal}
  {\bibinfo  {journal} {Phys. Rev. A}\ }\textbf {\bibinfo {volume} {64}},\
  \bibinfo {pages} {012310} (\bibinfo {year} {2001})}\BibitemShut {NoStop}%
\bibitem [{\citenamefont {Campagne-Ibarcq}\ \emph {et~al.}(2020)\citenamefont
  {Campagne-Ibarcq}, \citenamefont {Eickbusch}, \citenamefont {Touzard},
  \citenamefont {Zalys-Geller}, \citenamefont {Frattini}, \citenamefont
  {Sivak}, \citenamefont {Reinhold}, \citenamefont {Puri}, \citenamefont
  {Shankar}, \citenamefont {Schoelkopf}, \citenamefont {Frunzio}, \citenamefont
  {Mirrahimi},\ and\ \citenamefont {Devoret}}]{cqed_gkp}%
  \BibitemOpen
  \bibfield  {author} {\bibinfo {author} {\bibfnamefont {P.}~\bibnamefont
  {Campagne-Ibarcq}}, \bibinfo {author} {\bibfnamefont {A.}~\bibnamefont
  {Eickbusch}}, \bibinfo {author} {\bibfnamefont {S.}~\bibnamefont {Touzard}},
  \bibinfo {author} {\bibfnamefont {E.}~\bibnamefont {Zalys-Geller}}, \bibinfo
  {author} {\bibfnamefont {N.~E.}\ \bibnamefont {Frattini}}, \bibinfo {author}
  {\bibfnamefont {V.~V.}\ \bibnamefont {Sivak}}, \bibinfo {author}
  {\bibfnamefont {P.}~\bibnamefont {Reinhold}}, \bibinfo {author}
  {\bibfnamefont {S.}~\bibnamefont {Puri}}, \bibinfo {author} {\bibfnamefont
  {S.}~\bibnamefont {Shankar}}, \bibinfo {author} {\bibfnamefont {R.~J.}\
  \bibnamefont {Schoelkopf}}, \bibinfo {author} {\bibfnamefont
  {L.}~\bibnamefont {Frunzio}}, \bibinfo {author} {\bibfnamefont
  {M.}~\bibnamefont {Mirrahimi}},\ and\ \bibinfo {author} {\bibfnamefont
  {M.~H.}\ \bibnamefont {Devoret}},\ }\bibfield  {title} {\bibinfo {title}
  {Quantum error correction of a qubit encoded in grid states of an
  oscillator},\ }\href {https://doi.org/10.1038/s41586-020-2603-3} {\bibfield
  {journal} {\bibinfo  {journal} {Nature}\ }\textbf {\bibinfo {volume} {584}},\
  \bibinfo {pages} {368} (\bibinfo {year} {2020})}\BibitemShut {NoStop}%
\bibitem [{\citenamefont {Fl{\"u}hmann}\ \emph {et~al.}(2019)\citenamefont
  {Fl{\"u}hmann}, \citenamefont {Nguyen}, \citenamefont {Marinelli},
  \citenamefont {Negnevitsky}, \citenamefont {Mehta},\ and\ \citenamefont
  {Home}}]{ion_gkp}%
  \BibitemOpen
  \bibfield  {author} {\bibinfo {author} {\bibfnamefont {C.}~\bibnamefont
  {Fl{\"u}hmann}}, \bibinfo {author} {\bibfnamefont {T.~L.}\ \bibnamefont
  {Nguyen}}, \bibinfo {author} {\bibfnamefont {M.}~\bibnamefont {Marinelli}},
  \bibinfo {author} {\bibfnamefont {V.}~\bibnamefont {Negnevitsky}}, \bibinfo
  {author} {\bibfnamefont {K.}~\bibnamefont {Mehta}},\ and\ \bibinfo {author}
  {\bibfnamefont {J.~P.}\ \bibnamefont {Home}},\ }\bibfield  {title} {\bibinfo
  {title} {Encoding a qubit in a trapped-ion mechanical oscillator},\ }\href
  {https://doi.org/10.1038/s41586-019-0960-6} {\bibfield  {journal} {\bibinfo
  {journal} {Nature}\ }\textbf {\bibinfo {volume} {566}},\ \bibinfo {pages}
  {513} (\bibinfo {year} {2019})}\BibitemShut {NoStop}%
\bibitem [{\citenamefont {Nagata}\ \emph {et~al.}(2007)\citenamefont {Nagata},
  \citenamefont {Okamoto}, \citenamefont {O'Brien}, \citenamefont {Sasaki},\
  and\ \citenamefont {Takeuchi}}]{noon_optics}%
  \BibitemOpen
  \bibfield  {author} {\bibinfo {author} {\bibfnamefont {T.}~\bibnamefont
  {Nagata}}, \bibinfo {author} {\bibfnamefont {R.}~\bibnamefont {Okamoto}},
  \bibinfo {author} {\bibfnamefont {J.~L.}\ \bibnamefont {O'Brien}}, \bibinfo
  {author} {\bibfnamefont {K.}~\bibnamefont {Sasaki}},\ and\ \bibinfo {author}
  {\bibfnamefont {S.}~\bibnamefont {Takeuchi}},\ }\bibfield  {title} {\bibinfo
  {title} {Beating the standard quantum limit with four-entangled photons},\
  }\href {https://doi.org/10.1126/science.1138007} {\bibfield  {journal}
  {\bibinfo  {journal} {Science}\ }\textbf {\bibinfo {volume} {316}},\ \bibinfo
  {pages} {726} (\bibinfo {year} {2007})},\ \Eprint
  {https://arxiv.org/abs/https://www.science.org/doi/pdf/10.1126/science.1138007}
  {https://www.science.org/doi/pdf/10.1126/science.1138007} \BibitemShut
  {NoStop}%
\bibitem [{\citenamefont {Wolf}\ \emph {et~al.}(2019)\citenamefont {Wolf},
  \citenamefont {Shi}, \citenamefont {Heip}, \citenamefont {Gessner},
  \citenamefont {Pezz{\`e}}, \citenamefont {Smerzi}, \citenamefont {Schulte},
  \citenamefont {Hammerer},\ and\ \citenamefont
  {Schmidt}}]{fock_estimation_ion}%
  \BibitemOpen
  \bibfield  {author} {\bibinfo {author} {\bibfnamefont {F.}~\bibnamefont
  {Wolf}}, \bibinfo {author} {\bibfnamefont {C.}~\bibnamefont {Shi}}, \bibinfo
  {author} {\bibfnamefont {J.~C.}\ \bibnamefont {Heip}}, \bibinfo {author}
  {\bibfnamefont {M.}~\bibnamefont {Gessner}}, \bibinfo {author} {\bibfnamefont
  {L.}~\bibnamefont {Pezz{\`e}}}, \bibinfo {author} {\bibfnamefont
  {A.}~\bibnamefont {Smerzi}}, \bibinfo {author} {\bibfnamefont
  {M.}~\bibnamefont {Schulte}}, \bibinfo {author} {\bibfnamefont
  {K.}~\bibnamefont {Hammerer}},\ and\ \bibinfo {author} {\bibfnamefont
  {P.~O.}\ \bibnamefont {Schmidt}},\ }\bibfield  {title} {\bibinfo {title}
  {Motional fock states for quantum-enhanced amplitude and phase measurements
  with trapped ions},\ }\href {https://doi.org/10.1038/s41467-019-10576-4}
  {\bibfield  {journal} {\bibinfo  {journal} {Nature Communications}\ }\textbf
  {\bibinfo {volume} {10}},\ \bibinfo {pages} {2929} (\bibinfo {year}
  {2019})}\BibitemShut {NoStop}%
\bibitem [{\citenamefont {Lachman}\ and\ \citenamefont
  {Filip}(2022)}]{Radim_nongauss}%
  \BibitemOpen
  \bibfield  {author} {\bibinfo {author} {\bibfnamefont {L.}~\bibnamefont
  {Lachman}}\ and\ \bibinfo {author} {\bibfnamefont {R.}~\bibnamefont
  {Filip}},\ }\bibfield  {title} {\bibinfo {title} {Quantum non-gaussianity of
  light and atoms},\ }\href
  {https://doi.org/https://doi.org/10.1016/j.pquantelec.2022.100395} {\bibfield
   {journal} {\bibinfo  {journal} {Progress in Quantum Electronics}\ }\textbf
  {\bibinfo {volume} {83}},\ \bibinfo {pages} {100395} (\bibinfo {year}
  {2022})}\BibitemShut {NoStop}%
\bibitem [{\citenamefont {Hanamura}\ \emph {et~al.}(2021)\citenamefont
  {Hanamura}, \citenamefont {Asavanant}, \citenamefont {Fukui}, \citenamefont
  {Konno},\ and\ \citenamefont {Furusawa}}]{disp_estimation_theory}%
  \BibitemOpen
  \bibfield  {author} {\bibinfo {author} {\bibfnamefont {F.}~\bibnamefont
  {Hanamura}}, \bibinfo {author} {\bibfnamefont {W.}~\bibnamefont {Asavanant}},
  \bibinfo {author} {\bibfnamefont {K.}~\bibnamefont {Fukui}}, \bibinfo
  {author} {\bibfnamefont {S.}~\bibnamefont {Konno}},\ and\ \bibinfo {author}
  {\bibfnamefont {A.}~\bibnamefont {Furusawa}},\ }\bibfield  {title} {\bibinfo
  {title} {Estimation of gaussian random displacement using non-gaussian
  states},\ }\href {https://doi.org/10.1103/PhysRevA.104.062601} {\bibfield
  {journal} {\bibinfo  {journal} {Phys. Rev. A}\ }\textbf {\bibinfo {volume}
  {104}},\ \bibinfo {pages} {062601} (\bibinfo {year} {2021})}\BibitemShut
  {NoStop}%
\bibitem [{\citenamefont {Zurek}(2001)}]{subPlanck}%
  \BibitemOpen
  \bibfield  {author} {\bibinfo {author} {\bibfnamefont {W.~H.}\ \bibnamefont
  {Zurek}},\ }\bibfield  {title} {\bibinfo {title} {Sub-planck structure in
  phase space and its relevance for quantum decoherence},\ }\href
  {https://doi.org/10.1038/35089017} {\bibfield  {journal} {\bibinfo  {journal}
  {Nature}\ }\textbf {\bibinfo {volume} {412}},\ \bibinfo {pages} {712}
  (\bibinfo {year} {2001})}\BibitemShut {NoStop}%
\bibitem [{\citenamefont {Wu}\ and\ \citenamefont
  {Zhuang}(2021)}]{general_tscode}%
  \BibitemOpen
  \bibfield  {author} {\bibinfo {author} {\bibfnamefont {J.}~\bibnamefont
  {Wu}}\ and\ \bibinfo {author} {\bibfnamefont {Q.}~\bibnamefont {Zhuang}},\
  }\bibfield  {title} {\bibinfo {title} {Continuous-variable error correction
  for general gaussian noises},\ }\href
  {https://doi.org/10.1103/PhysRevApplied.15.034073} {\bibfield  {journal}
  {\bibinfo  {journal} {Phys. Rev. Appl.}\ }\textbf {\bibinfo {volume} {15}},\
  \bibinfo {pages} {034073} (\bibinfo {year} {2021})}\BibitemShut {NoStop}%
\bibitem [{\citenamefont {Kessler}\ \emph {et~al.}(2014)\citenamefont
  {Kessler}, \citenamefont {Lovchinsky}, \citenamefont {Sushkov},\ and\
  \citenamefont {Lukin}}]{qec_metrology}%
  \BibitemOpen
  \bibfield  {author} {\bibinfo {author} {\bibfnamefont {E.~M.}\ \bibnamefont
  {Kessler}}, \bibinfo {author} {\bibfnamefont {I.}~\bibnamefont {Lovchinsky}},
  \bibinfo {author} {\bibfnamefont {A.~O.}\ \bibnamefont {Sushkov}},\ and\
  \bibinfo {author} {\bibfnamefont {M.~D.}\ \bibnamefont {Lukin}},\ }\bibfield
  {title} {\bibinfo {title} {Quantum error correction for metrology},\ }\href
  {https://doi.org/10.1103/PhysRevLett.112.150802} {\bibfield  {journal}
  {\bibinfo  {journal} {Phys. Rev. Lett.}\ }\textbf {\bibinfo {volume} {112}},\
  \bibinfo {pages} {150802} (\bibinfo {year} {2014})}\BibitemShut {NoStop}%
\bibitem [{\citenamefont {Harrington}\ and\ \citenamefont
  {Preskill}(2001)}]{gaussian_quantum_channel}%
  \BibitemOpen
  \bibfield  {author} {\bibinfo {author} {\bibfnamefont {J.}~\bibnamefont
  {Harrington}}\ and\ \bibinfo {author} {\bibfnamefont {J.}~\bibnamefont
  {Preskill}},\ }\bibfield  {title} {\bibinfo {title} {Achievable rates for the
  gaussian quantum channel},\ }\href
  {https://doi.org/10.1103/PhysRevA.64.062301} {\bibfield  {journal} {\bibinfo
  {journal} {Phys. Rev. A}\ }\textbf {\bibinfo {volume} {64}},\ \bibinfo
  {pages} {062301} (\bibinfo {year} {2001})}\BibitemShut {NoStop}%
\bibitem [{\citenamefont {Miki}\ \emph {et~al.}(2017)\citenamefont {Miki},
  \citenamefont {Yabuno}, \citenamefont {Yamashita},\ and\ \citenamefont
  {Terai}}]{sspd}%
  \BibitemOpen
  \bibfield  {author} {\bibinfo {author} {\bibfnamefont {S.}~\bibnamefont
  {Miki}}, \bibinfo {author} {\bibfnamefont {M.}~\bibnamefont {Yabuno}},
  \bibinfo {author} {\bibfnamefont {T.}~\bibnamefont {Yamashita}},\ and\
  \bibinfo {author} {\bibfnamefont {H.}~\bibnamefont {Terai}},\ }\bibfield
  {title} {\bibinfo {title} {Stable, high-performance operation of a
  fiber-coupled superconducting nanowire avalanche photon detector},\ }\href
  {https://doi.org/10.1364/OE.25.006796} {\bibfield  {journal} {\bibinfo
  {journal} {Opt. Express}\ }\textbf {\bibinfo {volume} {25}},\ \bibinfo
  {pages} {6796} (\bibinfo {year} {2017})}\BibitemShut {NoStop}%
\bibitem [{\citenamefont {Noh}\ \emph {et~al.}(2020)\citenamefont {Noh},
  \citenamefont {Girvin},\ and\ \citenamefont {Jiang}}]{ts_code}%
  \BibitemOpen
  \bibfield  {author} {\bibinfo {author} {\bibfnamefont {K.}~\bibnamefont
  {Noh}}, \bibinfo {author} {\bibfnamefont {S.~M.}\ \bibnamefont {Girvin}},\
  and\ \bibinfo {author} {\bibfnamefont {L.}~\bibnamefont {Jiang}},\ }\bibfield
   {title} {\bibinfo {title} {Encoding an oscillator into many oscillators},\
  }\href {https://doi.org/10.1103/PhysRevLett.125.080503} {\bibfield  {journal}
  {\bibinfo  {journal} {Phys. Rev. Lett.}\ }\textbf {\bibinfo {volume} {125}},\
  \bibinfo {pages} {080503} (\bibinfo {year} {2020})}\BibitemShut {NoStop}%
\end{thebibliography}%
\end{document}